\documentclass{article}
\usepackage{graphicx}
\usepackage{amsmath,amsthm,amssymb,amscd}
\usepackage{authblk}

\begin{document}
\title{Afterpulsing model based on the quasi-continuous distribution of deep levels in single-photon avalanche diodes}
\author[1,2]{D.~B.~Horoshko\thanks{Email: dmitri.horoshko@univ-lille1.fr}}
\author[1]{V.~N.~Chizhevsky}
\author[1]{S.~Ya.~Kilin}
\affil[1]{B.~I.~Stepanov Institute of Physics, NASB, Nezavisimosti Ave.~68, Minsk 220072 Belarus}
\affil[2]{Univ. Lille, CNRS, UMR 8523 - PhLAM - Physique des Lasers Atomes et Mol\'ecules, F-59000 Lille, France}
\maketitle

\begin{abstract}
We have performed a statistical characterization of the effect of afterpulsing in a free-running silicon single-photon detector by measuring the distribution of afterpulse waiting times in response to pulsed illumination and fitting it by a sum of exponentials. We show that a high degree of goodness of fit can be obtained for 5 exponentials, but the physical meaning of estimated characteristic times is dubious. We show that a continuous limit of the sum of exponentials with a uniform density between the limiting times gives excellent fitting results in the full range of the detector response function. This means that in certain detectors the afterpulsing is caused by a continuous band of deep levels in the active area of the photodetector.
\end{abstract}

\section{Introduction}

Single photon detectors play a central role in modern quantum optics and quantum information science, providing a direct observation of single particles of light with high spatio-temporal resolution \cite{KilinRev,GisinRev}. The most widely used class of these detectors is represented by single-photon avalanche diodes (SPADs), operating in the so-called Geiger mode and producing a short pulse (photocount) in external electric chain as a result of single photon absorbtion \cite{ZappaRev}. The precision of quantum measurements is at present mainly limited by two fundamental imperfections of these highly sensitive devices: non-detection of some photons (non-unity quantum efficiency and dead-time) and production of counts caused by effects other than the absorbtion of a photon (dark counts and afterpulses). Though these imperfections cannot nowadays be overcome completely, in many applications, especially in ones related to security, like quantum cryptography and quantum random number generators, it is sufficient to provide a precise quantitative characterisation of the imperfections, thus enabling the users to distinguish the regular non-ideal detector operation from an intrusion of a malevolent adversary. And though at present the quantum efficiency, dead-time and dark count rate of SPADs can be measured with acceptable precision and are typically included into the detector specifications by the manufacturer, the situation with afterpulses is far less optimistic, notwithstanding considerable efforts in this area \cite{Zappa03,Jensen06,Itzler08,Gisin09,Weid11,Itzler12,Stipcevic13,Ursin14,Korzh14}. In particular the exact shape of the distribution function for the apterpulse waiting time is not generally known and the methods for its precise characterisation are still in the process of development. In the present work we develop an effective method for reliable determination of the parameters of this distribution and demonstrate its success in characterization of afterpulses in a commercially available silicon SPAD. As a result, we show that a model of continuous band of deep levels in the active area of the photodetector can give a very good description of the afterpulsing effect in the full range of waiting times, where afterpulsing is non-negligible, from tens of ns to tens of $\mu$s.

\section{Experimental setup}
\begin{figure}[ht]
\begin{center}
{\includegraphics[width=.8\columnwidth]{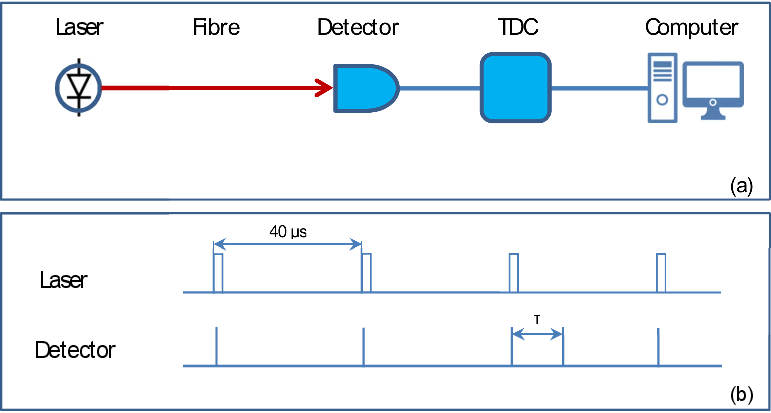}}
\caption{\label{Fig1}(a) Schematic of the experimental setup and (b) the temporal realisation of the laser pulses and the detector clicks. The recorded value of the afterpulse (or dark count) waiting time is shown as $\tau$.}
\end{center}
\end{figure}

Our setup is depicted in Fig. \ref{Fig1}a. The multimode-fibre coupled vertical cavity surface-emitting laser operating at 850 nm produces a sequence of pulses approximately 7.5 ns long with the repetition rate 25 kHz. These pulses pass through a variable optical attenuator (VOA) and illuminate the active area of a fibre-coupled single-photon detector (idQuantique id100-MMF50) built on the basis of a silicon SPAD. The sequence of detector counts is converted into a sequence of numbers (time intervals between successive counts) by a time to digital converter (TDC) having the time resolution (cycle duration) of 2.5 ns. The measured time intervals are transmitted to a personal computer (PC) and stored for later analysis.

The operation of the setup is as follows. The intensity of laser pulses is tuned by VOA to the minimal value at which there is still a photocount for each light pulse. Thus, the majority of measured time intervals corresponds to the duration of the repetition period (40 $\mu$s or $16000\pm2$ TDC cycles). Two photocounts cannot be produced by one laser pulse because the detector dead-time (45 ns) exceeds the pulse duration. In approximately 1\% of periods there will be registered a count not caused by light (see Fig. \ref{Fig1}b). This count can be produced either by thermal fluctuations (dark count) or by carriers trapped into the so-called deep levels in the SPAD active area during the preceding avalanche and released with the course of time (afterpulse) \cite{ZappaRev,Zappa03}. In principle, there can also be present some stray light of very low intensity, whose effect (stray light photocounts) is indistinguishable from the dark counts. For simplicity, these counts will also be considered as dark counts. All time intervals ending not at the period end, i.e. corresponding to dark counts and afterpulses, are extracted from the array of measured intervals by software and binned to produce a histogram. This histogram, upon normalization, approaches for large enough sample length the probability density function (pdf) of the afterpulse waiting time plus a constant responsible for the dark count rate. This pdf can be considered as detector response function to pulsed excitation and the problem of its precise modelling is the primary aim of the present article.

\section{Data fitting to a model}
\subsection{Fitting to a model of discrete deep levels}
The histogram obtained in the experiment, described in the previous section, contains no data for intervals less than the detector dead-time (18 bins or TDC cycles) and reaches its maximum not later than at bin 23. The increase between the bins 18 and 23 is explained by recovery of the detector after the dead-time period. A monotonically decreasing histogram is obtained by removing the first 22 bins. Its theoretical shape is given by the following consideration \cite{Zappa03}. Let us denote as $A(t)$ the population of a deep level and $A(0)$ its value 22 TDC cycles (55 ns) after an avalanche. Since the process of depopulation in solids is typically exponential, we write $A(t)=A(0)\exp{(-t/\tau)}$, where $\tau$ is the deep level lifetime. Then the probability of an avalanche in a short time interval $dt$ is $dP=-P_{aval}dA(t)=P_{aval}A_0\exp{(-t/\tau)dt/\tau}$, where $P_{aval}$ is the probability for a carrier in active area to produce an avalanche. For $N$ independent deep levels we obtain the following expression for pdf of afterpulses and dark counts, defined as  $p(t)=dP/dt$:
\begin{equation}\label{theory}
 p(t)=\sum_{i=1}^N\frac{u_i}{\tau_i}e^{-t/\tau_i}+v,
\end{equation}
where the subscript $i$ denotes the corresponding deep level, $u_i=P_{aval}A_i(0)$, and $v$ is the dark count rate.

The function defined by Eq.(\ref{theory}) for $N$ deep levels has $2N+1$ parameters which can be estimated by fitting the theoretical curve to the normalized histogram of experimental data. For this purpose we have used the method of nonlinear least squares in MATLAB 8.1. The results are summarized in Table 1. For this estimation 10 samples were recorded each having $5\times10^8$ intervals (total record time 60 hours). About $5.6\times10^6$ of afterpulses and dark counts were found in each sample, giving the probability of afterpulse or dark count in a period equal to $P_{ad}=0.0113$. For given number of deep levels $N$, taking values from 1 to 5, the parameters were estimated for each sample. The goodness of fit was measured by $R^2$ statistics, taking the maximal value of 1 for the perfect fit. The resulting 10 vectors of estimates were closely spaced except for one or two failures having much lower $R^2$. The average of all good fits is shown in the table. Determining the parameters of 6 and more exponentials was not possible by this method because of highly increased number of failures. Computing the $R^2$ statistics is traditional for the least-squares fitting, but for fits with high precision, where $R^2$ is very close to unity, one would like to know how much of the statistics fluctuation is due to the measurement noise and how much is due to the slight incorrectness of the model. It is known that $R^2$ follows a beta-distribution for the data with identical Gaussian noise \cite{DraperSmith}, but in a histogram of observed frequencies each bar has binomial distribution with its own mean and variance. Calculation of critical values for $R^2$ in this case seems to be problematic and we have chosen another way. An alternative estimation of the goodness of fit was done by calculating the $\chi^2$ statistics (with all bins as individual cells), which was normalized by its critical value $\chi^2_{crit}$ at 95\% confidence level. Normalized value of $\chi^2$ below unity means that the model curve coincides with the data up to statistical fluctuation for the given sample size \cite{Cramer}.

\begin{table}[h]
\centering
  \caption{\label{T1} Estimated relaxation times (ns) and the goodness-of-fit statistics for the model pdf, defined by Eq. (\ref{theory})}
    \begin{tabular}{cccccccc}
    \hline
    N & $\tau_1$ & $\tau_2$ & $\tau_3$ & $\tau_4$ & $\tau_5$ & $\chi^2/\chi^2_{crit}$ & $1-R^2$ \\
    \hline
    1 & 152.6 &  &  &  &  & 319 & 0.04\\
    2 & 86.25 & 730.0 &  &  &  & 34.5 & 0.003\\
    3 & 65.28 & 286.3 & 2080 &  &  & 3.57 & 4$\times10^{-4}$\\
    4 & 59.25 & 209.5 & 904.5 & 4295 &  & 1.23 & 2$\times10^{-4}$\\
    5 & 56.75 & 161.5 & 426.0 &  1516 &  6883 & 1.05 & 1.5$\times10^{-4}$\\
    \hline
    \end{tabular}
\end{table}

\begin{figure}[htbp]
\centerline{\includegraphics[width=0.8\columnwidth]{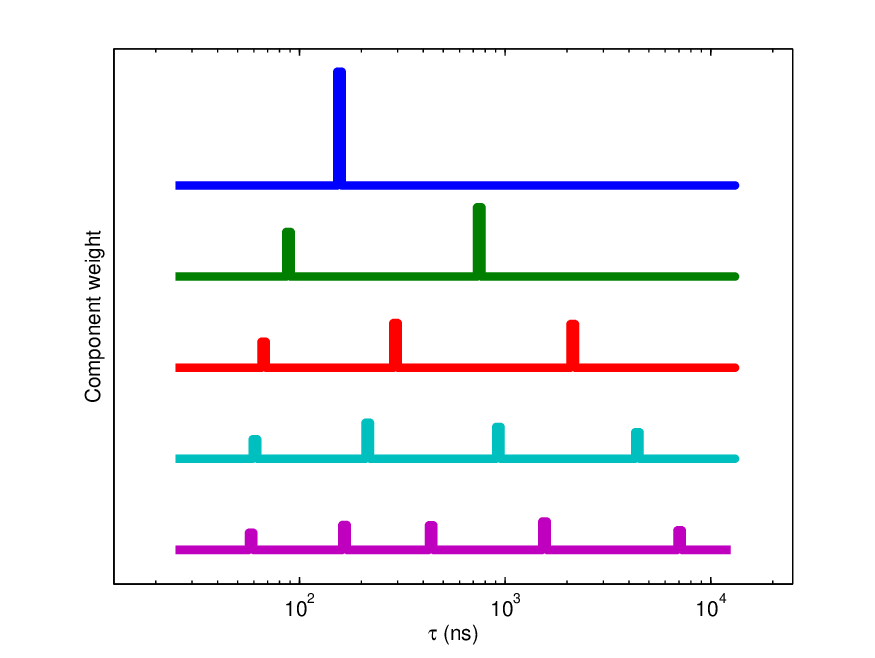}}
\caption{\label{Fig2}Diagram of characteristic times. Lines from top to bottom correspond to increasing number of deep levels in the model distribution Eq.(\ref{theory}) from $N=1$ to $N=5$. Vertical bars are placed at the estimated values of $\tau_i$, their heights give weights $u_i$ of the corresponding components.}%
\end{figure}

The analysis of Table \ref{T1} makes evident that the estimated values for $\tau_i$ hardly have physical meaning. Indeed, except for the shortest time, their values depend strongly on the hypothetic total number of deep levels $N$. The estimated times together with the corresponding weights $u_i$ are presented in a form of diagram in Fig. \ref{Fig2}.

\subsection{Fitting to a continuous band of deep levels}
Diagram in Fig. \ref{Fig2} has two interesting peculiarities. First, the estimated times seem to fill some range of times quasi-equidistantly in logarithmic time. Second, the weights seem to have rather close values for all components. This brings us to the idea that there may be \emph{a quasi-continuous distribution of deep levels with energies between some values $\epsilon_{min}$ and $\epsilon_{max}$ and constant density in $log(\tau)$ scale}.

In this case Eq.(\ref{theory}) should be replaced by
\begin{equation}\label{cont}
 p(t)=\int_{\epsilon_{min}}^{\epsilon_{max}}\gamma(\epsilon)e^{-\gamma(\epsilon)t}dU(\epsilon)+v,
\end{equation}
where $\gamma(\epsilon)=1/\tau(\epsilon)$ is the decay rate of the corresponding deep level and  $U(\epsilon)=P_{aval}A_{\epsilon}$,  where $A_{\epsilon}$ is the total population of deep levels with energies below or equal to $\epsilon$ just after the expiration of dead time. The decay rate of a deep level is known to obey the Arrhenius law \cite{Jensen06}:
\begin{equation}\label{Arrhenius}
\gamma(\epsilon)=Be^{\pm(\epsilon-\epsilon_0)/kT},
\end{equation}
where $B$, $\epsilon_0$ and the sign are parameters related to the carrier type and the band structure, $k$ is the Boltzmann constant and $T$ is the temperature. The sign in Eq.(\ref{Arrhenius}) is irrelevant for the shape of afterpulse distribution and for the rest of this article we chose the lower sign for definiteness. Our hypothesis of constant density in $log(\tau)$ scale means that
\begin{equation}\label{hypothesis}
dU(\epsilon)=Cd\left(log(\tau)\right)=-C{\gamma}^{-1}d\gamma=u_0d\epsilon,
\end{equation}
where $C$ and $u_0=C/kT$ are constants.

Substituting Eq.(\ref{hypothesis}) into Eq.(\ref{cont}) and integrating we easily obtain
\begin{equation}\label{sinhc}
 p(t)=2C\frac{\sinh\left(\Delta t\right)}{t}e^{-\gamma_0t}+v,
\end{equation}
where the parameters are $\gamma_0=\left(\gamma(\epsilon_{min})+\gamma(\epsilon_{max})\right)/2$ and $\Delta=\left(\gamma(\epsilon_{min})-\gamma(\epsilon_{max})\right)/2$. The first term in the r.h.s. of Eq.(\ref{sinhc}) represents an exponential multiplied by a function, which in analogy with the sinc function ${\rm sinc}(x)=\frac{\sin(x)}{x}$ is sometimes called `hyperbolic sinc' function ${\rm sinhc}(x)=\frac{\sinh(x)}x$.

The results of fitting the experimental data to the distribution Eq.(\ref{sinhc}) for 10 recorded samples are summarized in Table \ref{T2}. The limiting times are defined as $\tau_{min}=\tau(\epsilon_{min})$ and $\tau_{max}=\tau(\epsilon_{max})$. We see that the estimation procedure is reproducible, the relative fluctuation is 2.2\% for $\tau_{min}$ and 9.5\% for $\tau_{max}$.

\begin{table}[h]
  \centering
  \caption{\label{T2}Estimated limiting values of relaxation times (ns), the statistics and the estimated afterpulse probability $P_a$ (\%)}
    \begin{tabular}{cccccc}
    \hline
    Sample \# & $\tau_{min}$ & $\tau_{max}$ & $\chi^2/\chi^2_{crit}$ & $1-R^2$ & $P_a$ \\
    \hline
    1 & 17.17 & 1724 & 3.2 & $4\times10^{-4}$ & 0.87\\
    2 & 17.38 & 2062 & 1.9 & $3\times10^{-4}$ & 0.87\\
    3 & 17.33 & 2003 & 1.9 & $3\times10^{-4}$ & 0.87\\
    4 & 16.95 & 1729 & 3.6 & $4\times10^{-4}$ & 0.88\\
    5 & 16.94 & 1833 & 3.4 & $4\times10^{-4}$ & 0.87\\
    6 & 17.07 & 1991 & 2.4 & $3\times10^{-4}$ & 0.87\\
    7 & 17.02 & 2008 & 2.2 & $4\times10^{-4}$ & 0.87\\
    8 & 16.67 & 2015 & 2.6 & $3\times10^{-4}$ & 0.87\\
    9 & 16.52 & 2365 & 2.3 & $4\times10^{-4}$ & 0.89\\
    10 & 16.20 & 2076 & 2.5 & $4\times10^{-4}$ & 0.89\\
    \hline
    \end{tabular}
\end{table}

\begin{figure}[ht]
\centerline{\includegraphics[width=0.8\columnwidth]{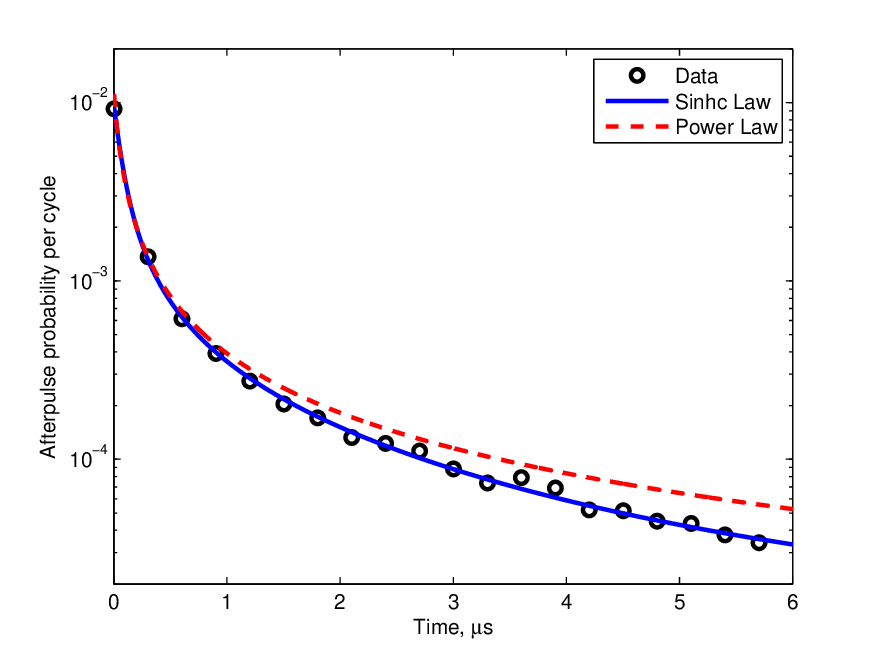}}
\caption{\label{Fig3}Observed distribution of afterpulses (one bin of every 120 shown) and its fits by the `hyperbolic sinc' function and by the power law $p(t)=0.41*(t+0.055)^{-1.15}$.}%
\end{figure}

Table \ref{T2} shows that the `hyperbolic sinc' distribution with only 2 parameters for afterpulsing gives a better goodness of fit than three exponentials in Eq.(\ref{theory}) having 5 independent parameters. This excellent fitting result is shown in Fig.\ref{Fig3}, where it is compared to another simple afterpulsing model: the power law, suggested recently for the description of afterpulse distribution in InGaAs/InP SPADs \cite{Itzler12}
\begin{equation}\label{power}
 p(t)=D/(t+t_d)^{\alpha}+v,
\end{equation}
where $D$ and $\alpha$ are positive parameters and $t_d$ is the dead time, 55 ns in our case. We have supplemented the power law of Ref.~\cite{Itzler12} by the dark count rate $v$, important for the description of long waiting times. For the power law the values of the statistics are $\chi^2/\chi^2_{crit}=11.5$ and $1-R^2=0.006$, which is much worse than that for the `hyperbolic sinc' distribution. It is seen in Fig. \ref{Fig3} that the power law, being very close to the data for short relaxation times, fails to explain the behavior for long relaxation times. Another drawback of the power law is very low (at the level of machine zero) value for the dark count rate $v$, when it is estimated together with the afterpulse distribution. The wide practice is to estimate the dark count rate separately and subtract it from the data. We prefer not to do so, since the application of $\chi^2$ statistics in this case is not justified.

Thus we see that a simple analytic form for the  afterpulse distribution function can be obtained from a simple assumption of constant density of populated levels per increment of energy, Eq.(\ref{hypothesis}). It should be noted that the relation Eq.(\ref{hypothesis}) was explored by Itzler and co-workers \cite{Itzler12} among other test densities for fitting the afterpulse distribution in InGaAs/InP SPADs \cite{Itzler12} and showed the results almost as good as the power law. In contrast, we came to Eq.(\ref{hypothesis}) by analysing the structure of the estimated characteristic times with growing number of exponentials in Eq.(\ref{theory}) and in our case (for Si SPADs) the `hyperbolic sinc' function shows better results when the entire detector response function is taken into account.

\begin{figure}[ht]
\centerline{\includegraphics[width=0.8\columnwidth]{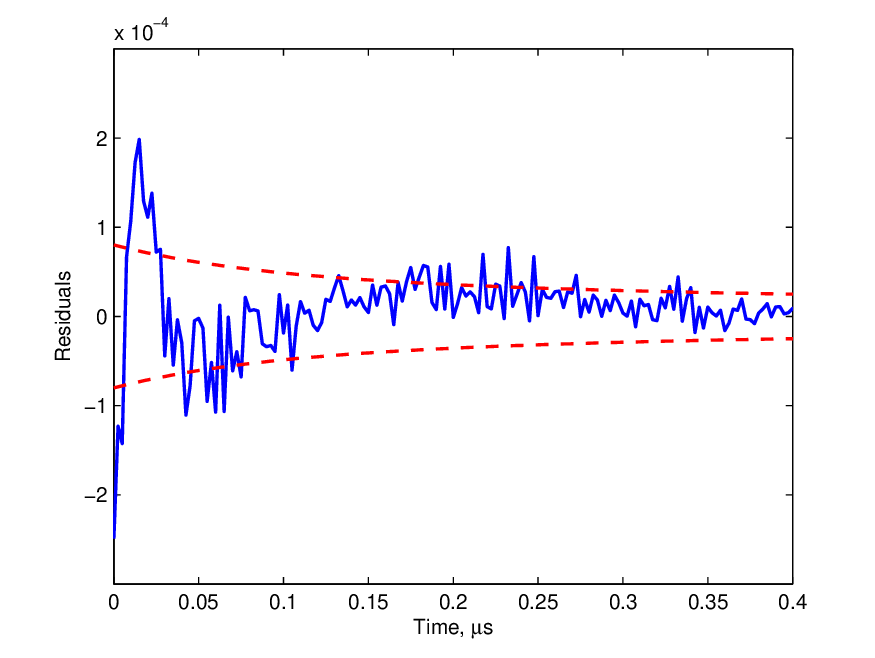}}
\caption{\label{Fig4}Residuals for the `hyperbolic sinc' model distribution given by Eq.(\ref{sinhc}) (solid line) and $\pm2\sigma$ limits for statistical fluctuations caused by finite sample size (dashed lines). Only short times are shown, the rest of the residuals are well within the $2\sigma$ region}%
\end{figure}

For better understanding the remaining imperfections of our model we plot in Fig. \ref{Fig4} the residuals, i.e. the differences between the observed and the model-predicted numbers of counts in each bin (normalized to the total count number $n$).

For a perfect fit about 95\% of the residuals are expected to lie in the limits $\pm2\sigma(t)$, where $\sigma(t)=\sqrt{p(t)\left(1-p(t)\right)/n}$. We see in Fig.~\ref{Fig4} that the residuals fall significantly outside the $2\sigma$ area for the region of very short times, which explains the above-unity value of normalized $\chi^2$ statistic. It may mean that for very short relaxation times our model requires some slight modifications, e.g. a more precise description of the SPAD recovery after the dead time period.

\subsection{Fitting the response of several detectors}
We have also applied the considered model to 3 other detectors of the same series (idQuantique id100-MMF50), being in our disposal. The results are summarized in Table \ref{T3}, where the fitting by the `hyperbolic sinc (sinhc) law', Eq.(\ref{sinhc}), is compared to other two models with three parameters: the exponential, given by Eq.(\ref{theory}) with $N=1$, and the `power law', given by Eq.(\ref{power}). We see that the `sinhc law' gives much lower values of the $\chi^2$ statistic, compared to other models. On the other hand, the values of statistic for detectors \# 2 and \# 4 even for the `sinhc law' are well above the critical value, which means that an adequate description of afterpulsing may require a more subtle model, combining a zone with one or several discrete levels. Such generalizations will be the subject of our further research in this area. For now we can state, that for the considered type of detectors on the basis of silicon SPADs, the zonal model of deep levels gives much better results than the alternative models.

\begin{table}[h]
  \caption{\label{T3}Normalized statistics $\chi^2/\chi^2_{crit}$ for 4 detectors, comparing 3 different models for the afterpulsing description with waiting times up to 40 $\mu s$. Data for each detector are averages over 10 samples of $5.6\times10^6$ afterpulses.}

  \begin{center}
    \begin{tabular}{cccc}
    \hline
    Detector \# & \hspace{3mm} exponential \hspace{3mm}& \hspace{3mm}power\hspace{3mm} & \hspace{3mm}sinhc\hspace{3mm} \\
    \hline
    1 & 319 & 732 & 2.88 \\
    2 & 909 & 1350 & 17.8 \\
    3 & 373 & 1640 & 1.48 \\
    4 & 333 & 872 & 15.8 \\
    \hline
    \end{tabular}
  \end{center}
\end{table}

\section{Conclusion}
In summary, we have shown that the finite sum of exponentials can be successfully used for fitting the afterpulse distribution in its full range, but the estimated parameters hardly have any physical meaning, since they vary significantly with the varying number of exponentials in the model function. Besides, we have shown that a continuous limit of the sum of exponentials with a natural assumption for density function of deep levels (constant per energy increment) leads to a simple analytical model function (Eq.(\ref{sinhc})) almost coincident with the data within the statistical error. We believe that at least in some cases two parameters of this model ($\tau_{min}$ and $\tau_{max}$) may be used for calibration of silicon SPADs instead of one `average afterpulsing time' giving rather loose description of the detector response function. We hope that the developed method of afterpulsing characterization will be helpful for security-oriented applications of quantum information science and to the precision measurements based on the photon counting technique.

\section*{Acknowledgements}
This work was funded by Belarus State Programme for Scientific Research `Convergence' and by Belarusian Republican Foundation for Fundamental Research.

\end{document}